# The Computational and Storage Potential of Volunteer Computing

David P. Anderson[1] and Gilles Fedak[2]

[1] Space Sciences Laboratory, U.C. Berkeley, davea@ssl.berkeley.edu
[2] INRIA, fedak@lri.fr

## Abstract

*"Volunteer computing" uses Internet-connected computers, volunteered by their owners, as a source of computing power and storage. This paper studies the potential capacity of volunteer computing. We analyzed measurements of over 330,000 hosts participating in a volunteer computing project. These measurements include processing power, memory, disk space, network throughput, host availability, user-specified limits on resource usage, and host churn. We show that volunteer computing can support applications that are significantly more data-intensive, or have high memory and storage requirements, than those in current projects.*

## 1. Introduction

**Volunteer computing** (also called "peer-to-peer computing" or "global computing") uses computers volunteered by the general public to do distributed scientific computing. Volunteer computing is being used in high-energy physics, molecular biology, medicine, astrophysics, climate study, and other areas. These projects have attained unprecedented computing power. For example, SETI@home has sustained a processing rate of about 60 TeraFLOPS for several years [3].

Most existing volunteer computing projects are throughput-oriented (i.e. they have minimal latency constraints), and have relatively small memory, disk, and network bandwidth requirements. To what extent is volunteer computing useful for more demanding applications? To explore this question, we studied the resources in the SETI@home host pool, and the various factors that limit their use.

We conclude that the potential of volunteer computing extends well beyond CPU-intensive tasks like SETI@home, and encompasses applications that require significant memory, disk space, and network throughput.

## 2. Resource measurements

SETI@home uses BOINC (Berkeley Open Infrastructure for Network Computing), a middleware system for volunteer computing [4]. BOINC facilitates the creation of volunteer computing projects; there are currently about 20 BOINC-based projects.

Volunteers participate by running a BOINC client program on their computers. They can attach each host to any set of projects, can control the resource share devoted to each project, and can limit when and how BOINC uses their computer resources.

The BOINC client periodically measures the hardware characteristics of the host. It also measures availability parameters such as the fraction of time the host is running and the fraction of time it has a network connection.

The BOINC client periodically contacts a scheduling server at each attached project, reporting the host's hardware and availability data. The scheduling server replies with a set of instructions for downloading executable files and input files, running the applications against the input files, and uploading the resulting output files.

This paper reflects the SETI@home host pool as of February 10, 2006, including only hosts that had successfully completed work within the past two weeks. Most of the data is available on the web at http://setiathome.berkeley.edu/stats/.

### 2.1. CPU performance

The BOINC client periodically executes the Whetstone [7] and Dhrystone [18] benchmarks. The results are interpreted as floating-point and integer operations per second, respectively.

Of the participating hosts, 25% have 2 or more CPUs, and 2% have 4 or more. A multiprocessor machine with N CPUs typically has lower performance than N times the speed of a single CPU. The difference is especially large for multi-core architectures, such as Intel "hyperthreaded" CPUs, which share a single floating-point unit between cores. To reflect this, BOINC benchmarks all CPUs simultaneously.

The CPU benchmark results are shown in Figures 1 and 2.

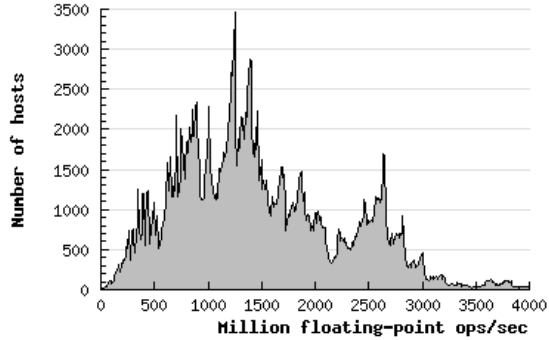

**Figure 1: Floating-point computing power**

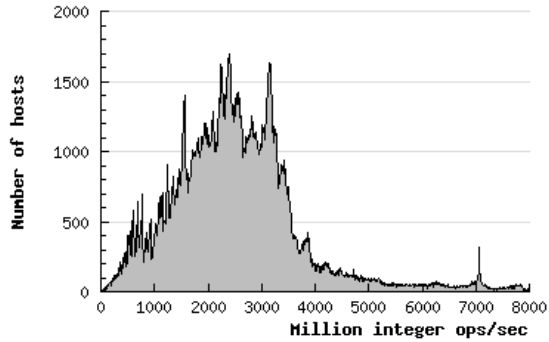

**Figure 2: Integer computing power**

The participating hosts run a variety of CPU types and operating systems, listed in Tables 1 and 2. Microsoft Windows, which accounts for 88.2% of the hosts and 91.6% of the FLOPS, is subdivided into versions.

| CPU type | Number of hosts | GFLOPS per host | GFLOPS total |
|---|---|---|---|
| Intel | 217,278 | 1.600 | 347,645 |
| AMD | 95,958 | 1.737 | 166,679 |
| PowerPC | 15,827 | 1.149 | 18,185 |
| SPARC | 1,035 | 0.755 | 781 |
| Others | 1,687 | 1.233 | 2,080 |
| Total | 331,785 | 1.613 | 535,169 |

**Table 1: CPU type breakdown**

| Operating system | Number of hosts | GFLOPS per host | GFLOPS total |
|---|---|---|---|
| Windows total | 292,688 | 1.676 | 490,545 |
| XP | 229,555 | 1.739 | 399,196 |
| 2000 | 42,830 | 1.310 | 56,107 |
| 2003 | 10,367 | 2.690 | 27,887 |
| 98 | 6,591 | 0.680 | 4,482 |
| Millennium | 1,973 | 0.789 | 1,557 |
| NT | 1,249 | 0.754 | 942 |
| Longhorn | 86 | 2.054 | 177 |
| 95 | 37 | 0.453 | 17 |
| Linux | 21,042 | 1.148 | 24,156 |
| Darwin | 15,830 | 1.150 | 18,205 |
| SunOS | 1,091 | 0.852 | 930 |
| Others | 1,134 | 1.364 | 1,547 |
| Total | 331,785 | 1.613 | 535,169 |

**Table 2: Operating system breakdown**

## 2.2. Memory

The BOINC client measures and reports the amount of physical memory (RAM) and swap space. Averages are 819 MB RAM and 2.03 GB swap. SETI@home uses about 32 MB of RAM.

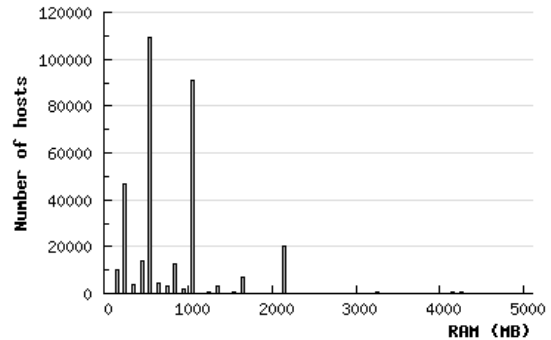

**Figure 3: RAM distribution**

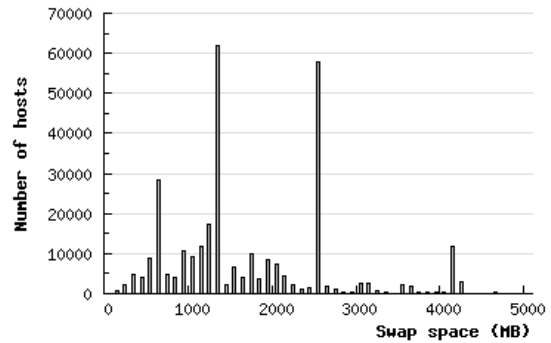

**Figure 4: Swap space distribution**

BOINC doesn't measure the usage of RAM or swap space by other applications.

### 2.3. Network throughput

The BOINC client measures throughput during periods when file transfers are in progress (many transfers may be active simultaneously) and maintains an exponentially weighted average of these values. These measurements reflect several factors: the network bandwidth between host and server, the speed with which the BOINC client transfers data, and the speed of the data server. We show only download throughput; SETI@home's upload files are too small to give meaningful data. The average throughput is 289 Kbps, and the distribution is shown in Figure 5.

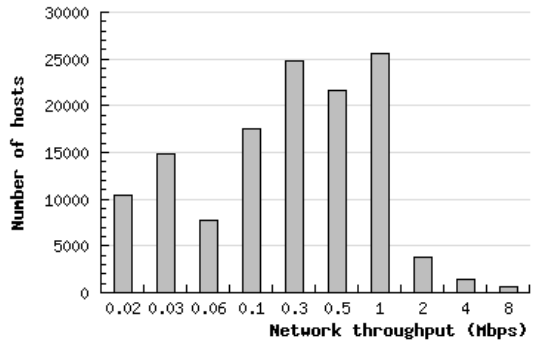

**Figure 5: Network download throughput distribution**

### 2.4. Disk space

The BOINC client measures the amount of total and free disk space on the volume where it is installed. Averages are 63 GB and 36 GB respectively (SETI@home uses about 10 MB per host). The total free space is 12.00 Petabytes. The distributions are shown in Figures 5 and 6.

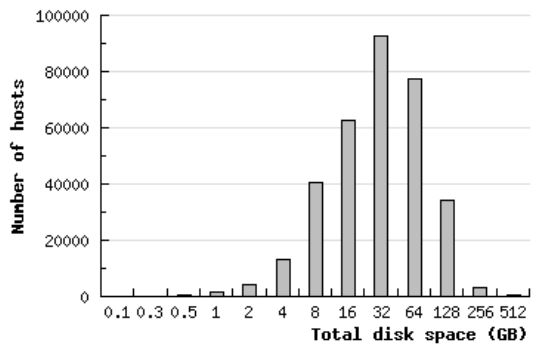

**Figure 6: Number of hosts versus total disk space**

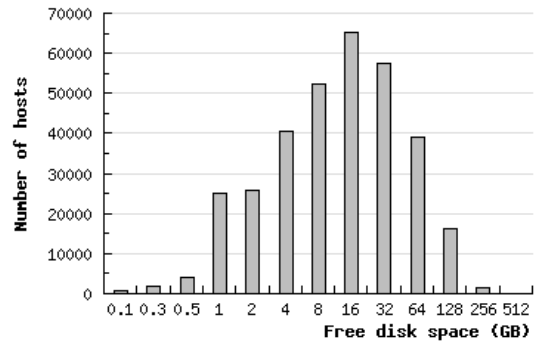

**Figure 7: Number of hosts versus free disk space**

BOINC doesn't measure space on volumes other than the one on which it is installed, so it may underestimate available disk space on some machines. It may overestimate disk space in situations where several hosts run BOINC from a shared network-accessible volume.

### 2.5. Combinations of resources

Hardware resources are meaningless in isolation. Disk space is useful only if there is network bandwidth available to access it, and CPU power is useful only if there is memory in which to execute. Figures 8 through 11 show various combinations of resources. Each graph shows the total amount of one resource given that the per-host amount of a second resource (shown on the X axis) exceeds a given value. Figures 8 to 10 are relevant to applications with large storage, memory, and network requirements respectively, while Figure 11 is relevant to applications involving data storage and retrieval.

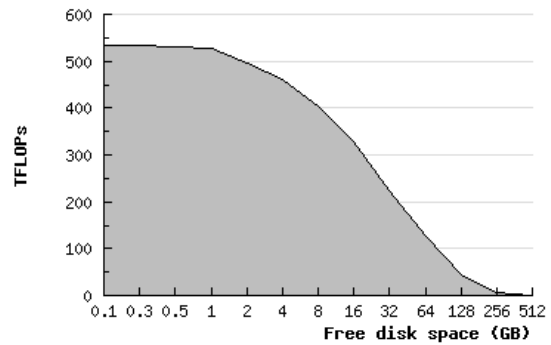

**Figure 8: Computing power versus free disk space**

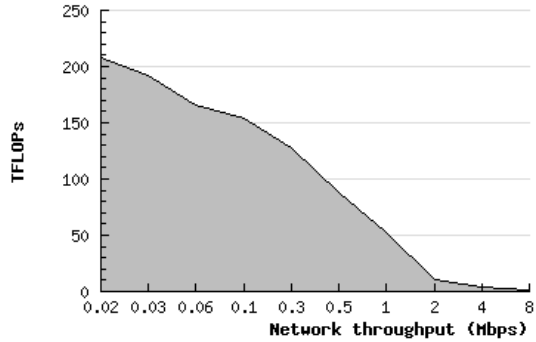

**Figure 9: Computing power versus network throughput**

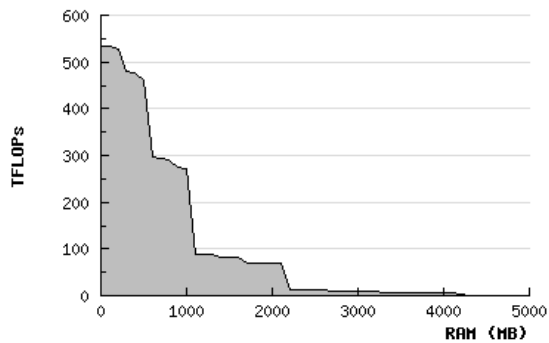

**Figure 10: Computing power versus memory size**

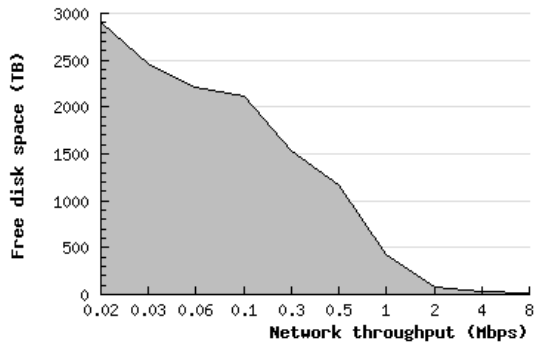

**Figure 11: Free disk space versus network throughput**

## 2.6. Host location

BOINC users, during the registration process, can specify their country. In this way hosts are associated with countries. The breakdown is shown in Table 3.

| Country | Number of hosts | GFLOPS per host | Disk free, GB | Thruput, Kbps |
|---|---|---|---|---|
| USA | 131,916 | 1.59 | 43.40 | 354.67 |
| Germany | 33,236 | 1.65 | 27.86 | 230.36 |
| UK | 23,638 | 1.62 | 40.37 | 297.59 |
| Canada | 14,821 | 1.54 | 38.00 | 449.82 |
| Japan | 12,931 | 1.49 | 36.76 | 266.58 |
| France | 9,412 | 1.76 | 29.52 | 212.86 |
| Australia | 7,747 | 1.60 | 34.38 | 298.10 |
| Italy | 6,921 | 1.73 | 31.17 | 206.45 |
| Netherlands | 6,609 | 1.66 | 28.36 | 226.61 |
| Spain | 6,418 | 1.59 | 30.29 | 168.98 |

**Table 3: Breakdown by country**

BOINC doesn't verify that users are actually from the country they indicate. However, the breakdown roughly agrees with time zone (offset from Greenwich Mean Time) reported by the BOINC client. The distribution of time zones is shown in Figure 12.

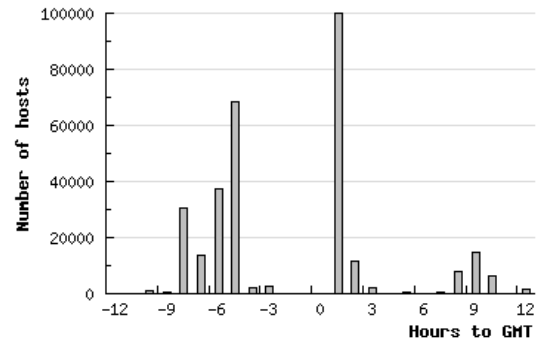

**Figure 12: Time zone distribution**

Users can specify whether hosts are at home, school, or work. We call this their **venue**. If users have multiple hosts, they can assign them different venues, and can define a different set of preferences (see Section 5) to each venue. For example, hosts at work might run BOINC applications only at night. The breakdown among venues is shown in Table 4.

| Venue | Number of hosts | GFLOPS per host | Disk free, GB | Through-put, Kbps |
|---|---|---|---|---|
| Home | 187,742 | 1.61 | 37.36 | 296.82 |
| Work | 52,484 | 1.73 | 32.00 | 411.87 |
| School | 12,023 | 1.62 | 32.19 | 344.37 |
| None | 79,535 | 1.54 | 35.47 | 198.32 |

**Table 4: Breakdown by venue**

## 3. Participation

### 3.1. Number of hosts

The dominant factor in a volunteer computing project's capacity is the number of participating hosts. This depends on many factors: the merit and public appeal of the application, the media coverage and other public relations activity, the incentives provided to users, and so on [6].

We expect that the number of hosts participating in volunteer computing will increase significantly, and that there will be many projects with hundreds of thousands of hosts. Currently, on the order of 1 million hosts participate – a few hundred thousand each for BOINC-based projects, GIMPS, distributed.net, Folding@home, Grid.org and World Community Grid. There are, according to current research, about 1 billion PCs in operation [9], so only about 0.1 percent of these participate. As volunteer projects appear in a wider range of areas, and are publicized and marketed more systematically, this percentage could increase by one or two orders of magnitude.

### 3.2. Host churn

A volunteer computing project's pool of hosts is dynamic: hosts continually arrive and leave. In addition, users occasionally reset the BOINC client on a given host, which has the effect of destroying one host and creating another.

We measured host "lifetime": the interval from creation to last communication for hosts that had not communicated in at least one month (this underestimates lifetime because it omits active hosts). The average host lifetime is 91 days, and the distribution is shown in Figure 13.

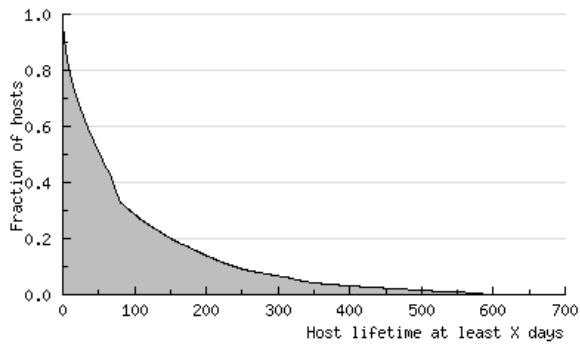

**Figure 13: Host lifetime distribution**

Host churn is important to applications that rely on their persistence on hosts. Examples include long-term storage applications like Oceanstore [14] and applications that do extremely long computations (such as Climateprediction.net, whose tasks take several months of CPU time on a typical host [6]).

The average number of active hosts is the average arrival rate times the average lifetime. The arrival rate can change over time. The arrival history for SETI@home is shown in Figure 14. Jumps in the graph correspond to public-relations events; gaps correspond to server outages.

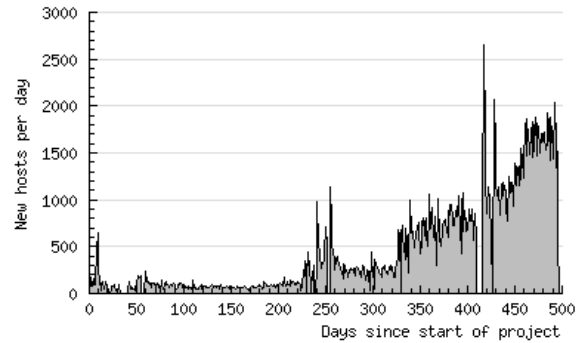

**Figure 14: Host arrival history**

### 3.3. Number of hosts per user

We analyzed the number of hosts per user (see Table 5 and Figure 15). The top two users had 2,987 and 1,783 hosts. Most users have a single host, but most hosts belong to a user with multiple hosts.

| Hosts per user | Number of users | Number of hosts | Percentage total hosts |
|---|---|---|---|
| 1 | 137,601 | 137,601 | 41.4% |
| 2-10 | 48,857 | 146,788 | 44.2% |
| 11-100 | 1,777 | 36,828 | 11.1% |
| 101-1000 | 30 | 5,799 | 1.7% |
| 1000+ | 2 | 4,770 | 1.4% |

**Table 5: Number of hosts per user**

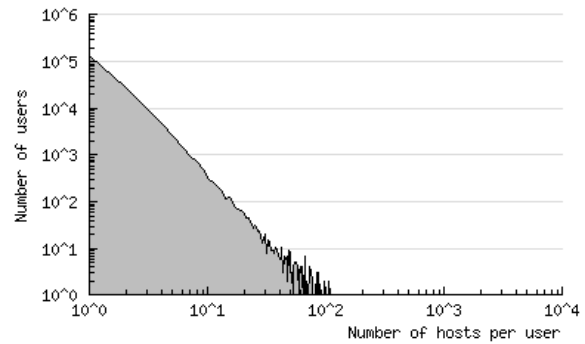

**Figure 15: Number of hosts per user**

## 4. Host availability

The BOINC client measures several aspects of host usage. The fraction of real time during which the BOINC client is running on the host is called the **on-fraction**. On most hosts, this is about the same as the fraction of time the host is powered on, since BOINC starts automatically at boot-up and runs in the background all the time. The mean on-fraction is 0.81.

The fraction (of the time that BOINC is running, not real time) that a physical network connection exists is called the **connected-fraction**. For hosts with LAN and DSL connections, this is close to 1. For hosts with telephone-based (ISDN or modem) or wireless connections, it may be lower. The mean connected-fraction is 0.83.

There may be periods when BOINC is running but is not allowed to execute applications or transfer files. This occurs when 1) the host is in use and user preferences are to not run when in use, 2) the time of day is outside a user-specified range, or 3) the user has explicitly suspended BOINC activity (via a command in the BOINC graphical interface). The fraction (of the time that BOINC is running, not real time) when BOINC is allowed to compute and communicate is called the **active-fraction**. The average active-fraction is 0.84.

Not all CPU time is available to BOINC: other CPU-intensive programs may run on some hosts. BOINC does not directly measure CPU load. Instead, it maintains, for each project, the **CPU efficiency**, defined as the average number of CPU seconds accumulated by that project's applications per second of real time during which they are runnable. This reflects CPU used for BOINC application graphics, CPU usage by non-BOINC applications, and I/O activity by the BOINC application.

In the case of SETI@home, which does very little I/O, CPU efficiency reflects primarily non-BOINC CPU load. The average CPU efficiency is 0.899.

## 5. User preferences

BOINC allows users to specify various **preferences** that limit how and when BOINC uses their resources. These preferences include:

**Run if user active**: whether BOINC should be active if there has been mouse or keyboard input in the last three minutes. The default is No, and 71.9% selected Yes.

**Active hours**: a range of hours during which BOINC may compute or communicate. 3.3% of users specified a range, with average duration 12.41 hours.

**Communication hours**: a range of hours during which BOINC may communicate. 0.8% of users specified a range, with average duration 12.18 hours.

**Confirm before connecting**: whether BOINC should get user permission before communicating. This is relevant to modem users and to low-latency applications. The default is No, and 8.4% selected Yes.

**Minimum connection interval**: a target minimum time between network connections. This has two purposes: 1) it lets modem users (who often pay a fee per connection) concentrate communication into infrequent bursts; 2) if a host (e.g. a laptop) is sporadically connected, the user can ensure that enough work is fetched to keep the host busy. The default is 0.1 days, and the average setting is 0.69 days.

**Disk access interval**: a minimum time between disk accesses. This is relevant to laptops with a low-power mode in which the disk turns off. The default is 60 seconds; the average setting is 78.9 seconds.

**Disk maximum used**: the maximum amount of disk space used by BOINC. The default is 100 GB. The average setting is 63.6 GB.

**Disk maximum percent used**: the maximum percentage of total disk space used by BOINC. The default is 50%. The average setting is 42.6%.

**Disk minimum free**: the minimum amount of free disk space. The default is 0.1 GB, and average setting is 0.97 GB.

In addition to these preferences, which apply to all projects to which a host is attached, users can specify a per-project **resource share** that determines how bottleneck resources are allocated. 16.8% of SETI@home users participate in other BOINC projects, and the average resource share of SETI@home (including those not participating in other projects) is 0.917.

## 6. Analysis

### 6.1. Total processing capacity

Because anonymously volunteered computers can't be trusted, many volunteer computing projects use redundant computing to minimize the effect of malicious or malfunctioning hosts. In this technique, each task is executed on two hosts belonging to different volunteers. If the results agree within application-defined tolerances, they are accepted; otherwise a third instance is executed, and so on. If the fraction of inconsistent results is low, redundant computing decreases effective computing power by a factor of slightly more than two.

Combining the factors we have presented in Sections 2, 3 4 and 5, and assuming that these factors are statistically independent, we have the following expression for the total floating-point computing power X available to a project:

$$X = X_{arrival} * X_{life} * X_{ncpus} * X_{flops} * X_{eff}$$
$$* X_{onfrac} * X_{active} * X_{redundancy} * X_{share}$$

Where $X_{arrival}$ is the average arrival rate of hosts, $X_{life}$ is the average lifetime of hosts, $X_{ncpus}$ is the average number of CPUs per host, $X_{flops}$ is the average FLOPS per CPU, $X_{eff}$ is the average CPU efficiency, $X_{onfrac}$ is the average on-fraction, $X_{active}$ is the average active-fraction, $X_{redundancy}$ is the reciprocal of the average redundancy, and $X_{share}$ is the average resource share (relative to other CPU-intensive projects).

For applications that use large amounts of RAM or disk, this estimate must be scaled by the factors described in sections 2.5 and 6.2. Analogous expressions estimate the limits of storage capacity and network transfer.

In the case of SETI@home, the product of the first four factors (i.e. the hardware resource) is about 535 TeraFLOPS. The product of the remaining five factors is 0.28. Thus SETI@home, at the time of this study, had a potential processing rate of 149.8 TeraFLOPS.

## 6.2. Data-intensive applications

To what extent can volunteer computing handle data-intensive tasks (i.e. those with large input files)? Foster and Iamnitchi discuss this question [10], and point out that while SETI@home processes about 25 KB of data per CPU hour, some applications have a much higher ratio. They cite astrophysics applications that process 60 MB and 660 MB per CPU hour.

To study this question, we define the **data rate** R of an application to be the average number of Mbytes of data it processes using 3.6e12 floating-point operations (i.e. one hour of CPU time on a 1 GFLOPS computer). We assume that a client is able to do both computation and communication nearly all the time (the BOINC client overlaps these activities, and network communication takes little CPU time).

Suppose a 1 GFLOPS computer has a 1 Mbps network connection. Then it can download 450 MB per hour. If it runs an application for which R=450, both network and CPU are saturated (i.e. busy all the time). If R < 450, the network is not saturated; if R > 450, the CPU is not saturated (of course, the excess CPU time could be used by a less data-intensive project).

This critical value of R varies with the host; it will be smaller if the host has a faster CPU or a slower network connection. For a given value of R, some hosts will be network-saturated and won't be able to devote all their CPU time to the application. Figure 15 illustrates this effect, showing the computing power available as a function of R. The shaded line shows the fraction of hosts whose CPUs are not saturated at the given data rate.

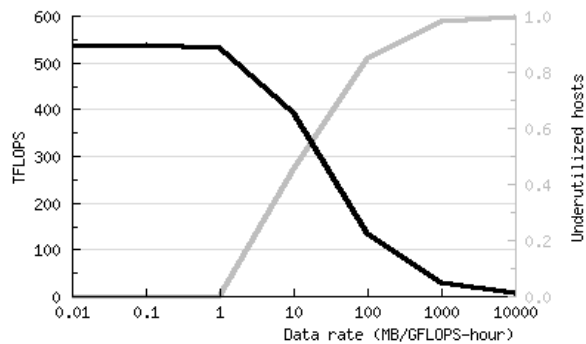

**Figure 15: Computing power versus data rate**

It can be seen that considerable processing power (tens or hundreds of TeraFLOPS) is available even to applications with R = 100 or 1,000. Thus volunteer computing can potentially handle data-intensive applications.

This analysis omits some important factors: saturating lots of client network connections could swamp the outgoing server links, ISP backbone networks, and shared incoming links. Solving these problems raises numerous research issues; we believe that an approach based on gleaning unused network bandwidth could be effective.

## 7. Related work

Sarmenta [17] articulated the idea of volunteer computing, and explored many of its technical issues. The Entropia [5] and XtremWeb [11] projects studied the speedup of specific applications in the context of volunteer computing. Gray [12] analyzed the economics of volunteer computing.

The Condor project [15] pioneered using the idle time of organizational workstations to do parallel computing. Other projects have studied the statistics of host availability [2, 13, 19]. Acharya and Setia [1] studied the availability of idle RAM on workstation pools. Eggert and Touch [8] studied operating system mechanisms for efficient use of idle resources.

Workstation cycle-stealing (and Grid computing in general) differs fundamentally from volunteer computing. It generally requires that parallel tasks run simultaneously, so that they may communicate; this in turn requires the ability to migrate running tasks. Resources are trusted, so that validation techniques like redundant computing are not needed. Workstations can be contacted dynamically (in BOINC, all communication is client-initiated, so that firewalls and NATs can be traversed).

## 8. Conclusion

We have analyzed the hardware characteristics of the hosts participating in a typical volunteer computing project, and have described various factors that affect the computing power and storage capacity available to the project. The host pool provides processing at a sustained rate of 95.5 TFLOPS. We have shown that it can provide lesser but still significant processing power for data-intensive applications. It also has the potential to provide 7.74 Petabytes of storage, with an access rate of 5.27 Terabytes per second.

We have provided a variety of data about host type and location. This data can be used to help volunteer computing projects decide what platforms to support and how to recruit participants.

In the future we plan to extend BOINC to allow peer-to-peer communication, as this will increase its capacity for applications with large intermediate files or replicated input files. This will require knowledge of peer-to-peer connectivity and bandwidth; we may use an existing system such as DIMES [17] for this purpose. We also plan to use BOINC data to study the change in Internet resources over time. We currently have about 10 months of historical host information, but the rapid change in the host pool makes it hard to derive meaningful conclusions from this data.

We thank Rom Walton, Matt Lebofsky, and many volunteer programmers for their help in collecting performance data, and we thank several colleagues who read and commented on the paper. This work was supported by the National Science Foundation grants SCI-0221529 and SCI-0438443.